\documentclass[12pt]{iopart} 
\usepackage{amstext}
\usepackage{iopams}  
\usepackage{geometry}
\usepackage{cite}
\usepackage{graphicx}
\usepackage{amssymb}

\usepackage{float}

\usepackage{url}
\usepackage{tikz}

\newcommand{\ket}{\rangle}
\newcommand{\subcaptionOverlay}[3]{
  \begin{tikzpicture}
    \node [inner sep=0,anchor=north west]at (#3) (image) {#1};
    \draw node [black] {#2};
  \end{tikzpicture}
}

\begin{document}

\title{Fast optical transport of ultracold molecules over long distances}

\author{Yicheng Bao$^{1,2}$, Scarlett S. Yu$^{1,2}$, Lo{\"\i}c Anderegg$^{1,2}$, Sean Burchesky$^{1,2}$, Derick Gonzalez-Acevedo$^{1,2}$, Eunmi Chae$^{3}$, Wolfgang Ketterle$^{2,4}$, Kang-Kuen Ni$^{1,2,5}$, John M. Doyle$^{1,2}$}

\address{
$^{1}$Department of Physics, Harvard University, Cambridge, MA 02138, USA\\
$^{2}$Harvard-MIT Center for Ultracold Atoms, Cambridge, MA 02138, USA\\
$^{3}$Department of Physics, Korea University, Seongbuk-gu, Seoul, 02841, South Korea\\
$^{4}$Department of Physics, Massachusetts Institute of Technology, Cambridge, MA 02139, USA\\
$^{5}$Department of Chemistry and Chemical Biology, Harvard University, Cambridge, MA 02138, USA
}

\ead{bao@g.harvard.edu}

\vspace{10pt}
\begin{abstract}
Optically trapped laser-cooled polar molecules hold promise for new science and technology in quantum information and quantum simulation. Large numerical aperture optical access and long trap lifetimes are needed for many studies, but these requirements are challenging to achieve in a magneto-optical trap (MOT) vacuum chamber that is connected to a cryogenic buffer gas beam source, as is the case for all molecule laser cooling experiments so far. Long distance transport of molecules greatly eases fulfilling these requirements as molecules are placed into a region separate from the MOT chamber. We realize a fast transport method for ultracold molecules based on an electronically focus-tunable lens combined with an optical lattice. The high transport speed is achieved by the 1D red-detuned optical lattice, which is generated by interference of a focus-tunable laser beam and a focus-fixed laser beam. Efficiency of $48(8)\%$ is realized in the transport of ultracold calcium monofluoride (CaF) molecules over $46\,\text{cm}$ distance in $50\,\text{ms}$, with a moderate heating from $32(2)\,\mu\text{K}$ to $53(4)\,\mu\text{K}$. Positional stability of the molecular cloud allows for stable loading of an optical tweezer array with single molecules.
\end{abstract}

\section{Introduction}
Ultracold molecules have a wide range of potential applications from quantum simulation and computation to searches for beyond the Standard Model particles \cite{demille2002quantum,bohn2017cold,carr2009cold,safronova2018search,kozyryev2017precision,kozyryev2021enhanced,altman2021quantum}. Rich internal states, long coherence times, and tunable dipolar interactions between ultracold polar molecules could be harnessed for quantum information storage and gate operations \cite{ni2018dipolar,hughes2020robust,gregory2021robust,caldwell2020enhancing,caldwell2021general}. Direct laser cooling of molecules has recently realized magnetic and optical trapping of molecules \cite{mccarron2018magnetic,williams2018magnetic,anderegg2018laser,langin2021polarization}, optical tweezer arrays of molecules \cite{anderegg2019optical}, the determination of rotational qubit coherence times \cite{burchesky2021rotational}, and the laser cooling of polyatomic molecules \cite{kozyryev2017sisyphus,baum20201d,augenbraun2020laser,vilas2021magneto}. Future advances will demand long trap lifetimes in an ultra-high vacuum (UHV) ``science chamber" with high numerical aperture optical access, which allows for both high fidelity imaging and strong dipole coupling of molecules for quantum gate operations. Another concurrent desire is high experimental repetition rates, leading to efficient data collection and higher statistical precision in all of the envisioned new science with ultracold molecules. Laser cooling of molecules starts from a cryogenic buffer gas beam (CBGB) source, where residual helium can enter the UHV chamber and potentially reduce the lifetime of molecules in the trap. Although molecular lifetimes much greater than $1\,\text{s}$ are routinely achieved, a separate science chamber can offer even longer lifetimes and, in addition, greater flexibility for future upgrades such as use of in situ high voltage electrodes and cryogenic blackbody radiation shields. All of these desired features could be realized with long distance transport to shuttle trapped molecules rapidly into a clean science chamber with large optical access.

Several challenges arise when transporting laser-cooled molecules. Vibrational excitation by room temperature blackbody radiation can pose a limit on the trap lifetime of polar molecules \cite{buhmann2008surface,williams2018magnetic}. This implies that the transport should be much faster than the blackbody lifetime. Furthermore, typical molecular laser cooling experiments have an experimental cycle time of less than a second, this short cycle time with fast transport is highly desired. Although various transport schemes for cold atoms have already been demonstrated, most of these methods cannot provide both the high trap depths and rapid transport speeds needed for laser-cooled molecules. The two main methods for atoms are magnetic and optical moving trap transport. Magnetic transport can be realized by mechanically translating a magnetic coil \cite{lewandowski2003simplified}, or controlling the sequence of current into a set of coils placed along the transport path \cite{greiner2001magnetic,minniberger2014magnetic}. Magnetic transport typically produces a low density sample in a large volume trap, and can conflict with optimal optical access, while optical transport can be implemented with minimum obstruction of optical access. There are many different optical transport schemes based on moving the focus of an optical dipole trap (ODT), including using a focus-tunable lens \cite{leonard2014optical}, rotating a Moir\'e pattern lens \cite{unnikrishnan2021long}, or, even simpler, mechanically moving a fixed lens on a translation stage \cite{gustavson2001transport,gross2016all,naides2013trapping}. These methods are often speed limited ultimately by the low axial trap frequency of the ODT. There are also demonstrations of using moving 1D optical lattice to shuttle atoms \cite{schmid2006long,middelmann2012long,klostermann2021fast}, where one beam is a zero-order Bessel beam to maintain constant waist size over long distance.

Here we present a hybrid transport approach based on the combination of electrically tunable lens (moving ODT focus) and a 1D red-detuned optical lattice. This approach offers a substantially higher transport velocity than a simple moving ODT transport, due to the high trapping frequency of the 1D lattice direction. The ODT in this approach uses technically moderate laser power while still maintaining a trap that is deep enough for directly loading laser-cooled molecules from a molasses. Using this hybrid approach, optically trapped CaF molecules are transported one way with an efficiency of $48(8)\%$ over $46\,\text{cm}$ in $50\,\text{ms}$. The short transport time results in negligible blackbody radiation loss, as well as having the advantage of high experimental repetition rate.

\section{Experiment setup}

\begin{figure}[!htbp]
\includegraphics[width=\columnwidth]{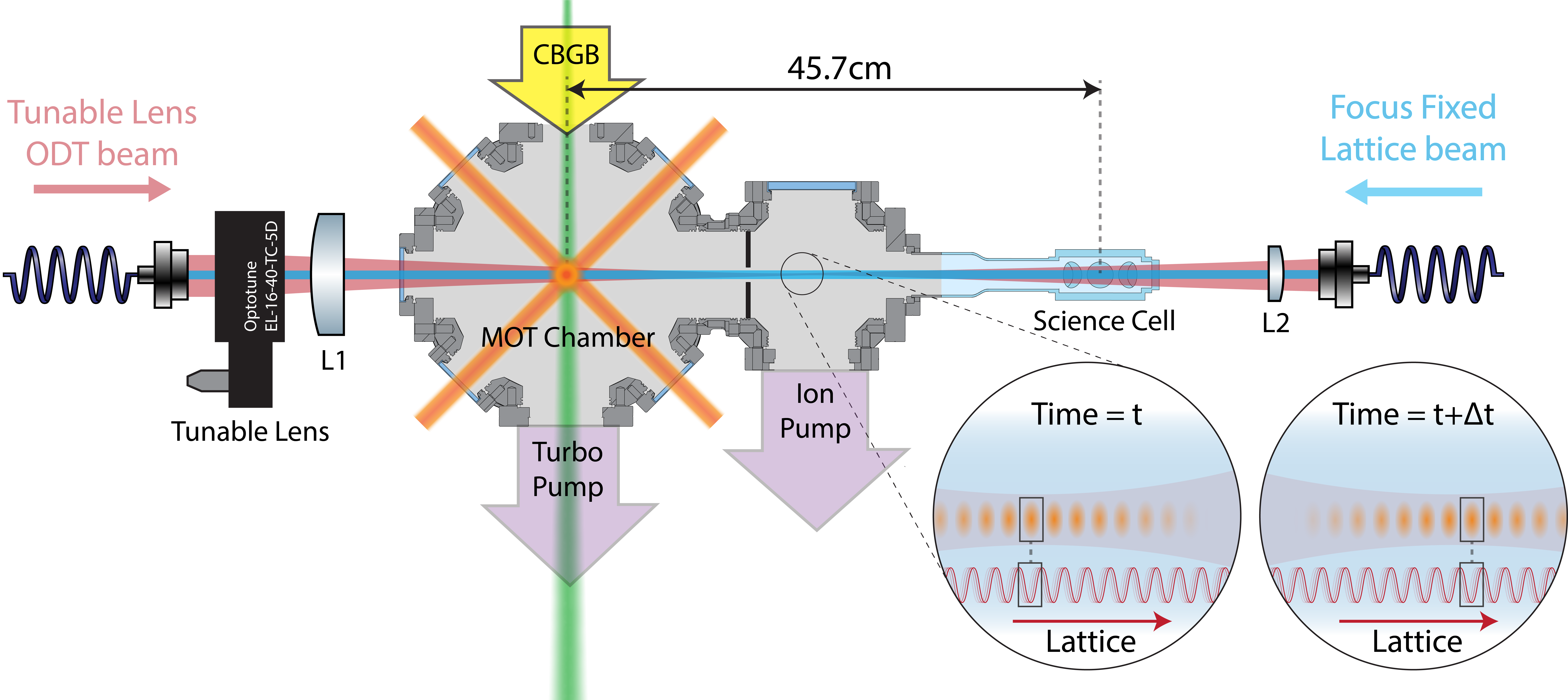}
\caption{Experiment setup. The yellow arrow indicates the direction of molecular beam from the cryogenic buffer gas beam (CBGB) source. Purple arrows denote the vacuum pumps to remove helium buffer gas from the chamber. The orange beams are the MOT beams. The green beam is the slowing beam. The blue beam is the large waist lattice beam, propagating from science cell to MOT chamber direction. The pink beam is the focus-tunable small waist ODT beam, propagating from MOT chamber to science cell direction. The thick black line between MOT chamber and pumping manifold is the differential pumping aperture. The inset shows a configuration in which the lattice is moving towards the science cell, while the tunable lens focus tracks the position of the molecules during one way transport. The beam diameter and divergence are not scaled and are for illustration only.}
\label{fig:setup}
\end{figure}

Our UHV system consists of a MOT chamber and a science cell, separated by a pumping manifold, shown in \Fref{fig:setup}. The distance between the centers of the MOT chamber and the science cell is $\sim46\,\text{cm}$. A $1.2\,\text{cm}$ diameter differential pumping aperture between the pumping manifold and MOT chamber reduces the helium conductance into the science cell. The science cell is an octagonal glass cell with two 3-inch diameter primary viewports. This optical access allows for two high numerical aperture microscope objectives to be used for projecting tightly focused optical tweezers and high fidelity imaging of single molecules.

\subsection{Transport lattice setup}

To obtain the largest trap depth in a lattice with a fixed amount of laser power, it is optimal to match the intensity of lattice beams propagating in the opposite direction, as this will exploit the full contrast of the interference. In order to achieve such intensity balance in a moving lattice over tens of centimeters for transport, both beams need to be focus tunable. However, this configuration is sensitive to misalignment and various drifts. In this work, we instead choose to use a large waist gaussian beam ($w=320\,\mu\text{m}$) with its focus fixed at the center of the transport path. The trap depth provided solely by this beam is less than $7\,\mu\text{K}$, too shallow to hold molecules on its own. To form a lattice, we counter-propagate a focus-tunable beam (ODT beam), which is a small waist ($w=57\,\mu\text{m}$) gaussian beam generated by a tunable lens. This tight ODT beam itself allows for a deep trap (typically $157\,\mu\text{K}$), which facilitates simple and efficient loading of CaF molecules from $\Lambda$-enhanced gray molasses. The ODT also holds molecules against gravity during the transport. In presence of both beams ($\lambda=1064\,\text{nm}$), the interference between them creates a 1D lattice with $532\,\text{nm}$ spacing between lattice sites. This 1D lattice has a peak-to-trough potential of $\sim100\,\mu\text{K}$ at the center position of the MOT chamber, sufficient to trap molecules at $32\,\mu\text{K}$. The axial trap frequency in the 1D lattice at the typical $100\,\mu\text{K}$ trap depth is estimated to be $157\,\text{kHz}$. Without the large waist focus-fixed beam, the axial trap frequency in the ODT would be only $\sim3.5\,\text{Hz}$, much lower than that of the 1D lattice and cannot support high acceleration needed for a fast transport.

The velocity of the moving lattice sites is proportional to the laser frequency difference of the two beams, expressed as $v=\frac{\lambda}{2}(f_{\text{lattice}}-f_{\text{ODT}})$. The final position of the transported molecules is controlled by the motion of the lattice sites, which is defined by the total phase difference accumulated during frequency sweeps, rather than the focus of the tunable lens. This greatly improves the stability of the final position against various drifts in tunable lenses \cite{leonard2014optical}. The tunable lens used in our setup is not temperature controlled nor is feedback used, and stable transport is still achieved.

To reach transport velocities in the tens of $\text{m/s}$ range, the frequencies of the two beams require several MHz detuning while maintaining phase coherence. We generate these two laser beams using two high power single-frequency polarization-maintaining Ytterbium doped fiber amplifiers seeded by a low noise diode laser. Relative frequency shifting between two high power lasers is achieved with a wide-band acousto-optic deflector (AOD) in a double-pass configuration. The center frequency of the AOD is $80\,\text{MHz}$ and a bandwidth of $\pm20\,\text{MHz}$ can be swept with reasonable diffraction efficiency. This allows for a maximum relative detuning of $\pm2\times20\,\text{MHz}$ between the two beams for round trip transport, corresponding to a maximum transport velocity of $21\,\text{m/s}$ in both directions. The seed laser is shifted by a double passed AOD before being amplified by one of the fiber amplifier for the lattice beam. A beam split from the seed laser is fed into a second fiber amplifier to create the ODT beam. These beams are coupled through fibers and deliver 30W for each beam onto the experimental table. 

The tunable lens\footnote{EL-16-40-TC-NIR-5D from Optotune} is chosen for its ability to tune between convex and concave lens shapes with a small wavefront error ($<0.15\,\lambda$ RMS over $16\,\text{mm}$ aperture), as well as good passive thermal stability and relative large aperture. To avoid coma aberration originated from gravity dragging the soft membrane in tunable lenses, the lens is installed with its optical axis aligned to gravity. To maintain a constant trap depth at the focus during the transport, the tunable lens is placed at the focal plane of the main focusing lens L1. With a collimated laser beam incident on the tunable lens, the trap focus after L1 can be moved along the optical axis without changing the waist size at the focus. The total optical path length difference caused by tuning the lens is $\sim100\,\lambda$ at $\lambda=1064\,\text{nm}$, which is much smaller than the motion generated by lattice frequency sweeping.

The tunable lens has mechanical resonances at frequencies below 1kHz. Excitation of higher order mechanical modes of the lens membrane can distort the wavefront of the laser beam, causing large fluctuations of the trap depth and the focal position. This can lead to rapid heating and loss of molecules, limiting the practical tuning speed of the tunable lens. This is typically not a problem in a tunable-lens only moving ODT transport system since the transport takes order of seconds, which is more than an order of magnitude slower than our scheme. To maintain a good beam quality while still forcing the lens to respond as fast as possible, the lens is driven with an optimized waveform instead of a simple linear ramp waveform. The optimized driving waveform is obtained by measuring the step function response of the lens, which is then inputted into a convex optimization algorithm to generate a waveform for the target focus trajectory (See supplemental material for details) \cite{iwai2019speeded,cvx,gb08}. Assuming the lens can be modeled as a linear time-invariant system, we linearly scale the optimized waveform to control the start and stop position of the lens without generating the optimized waveform each time.

The lattice beam is generated by sending a collimated beam into the $f=1000\,\text{mm}$ lens L2. The focus is placed at the middle of the transport path to reduce variation of lattice depth during the transport. The large waist also eases the alignment process of the two beams.

\subsection{Experimental sequence}

Our experiment begins with CaF molecules produced in a CBGB and then loaded into a 3D radio-frequency magneto-optical trap (RF MOT) operating on the $|X ^2\Sigma^+,N=1\ket \rightarrow |A ^2\Pi_{1/2},J=1/2\ket$ cycling transition \cite{barry2014magneto,mccarron2015improved, norrgard2016submillikelvin,anderegg2017radio}. The RF MOT is initially loaded at low magnetic field gradient and ramped up to compress the cloud to Gaussian RMS width of $0.8\,\text{mm}$. The RF magnetic field is then turned off and six MOT laser beams (now with polarization switching turned off) are blue detuned 30MHz to the X-A transition with only F=2 and F=1- components left on to perform $\Lambda$-enhanced gray molasses cooling \cite{anderegg2018laser,cheuk2018lambda,caldwell2019deep,ding2020sub,wu2021high}. The molecular cloud is quickly cooled to a temperature of $10\,\mu\text{K}$ in free space. The lattice beam and ODT beam are then turned on simultaneously. With $\Lambda$-enhanced gray molasses also working in the optical dipole potential, molecules are efficiently loaded into the transport lattice.

At the start of transport, the relative frequency between the ODT beam and lattice beam is linearly ramped up to $f_{\text{max}}$ (typically $18\,\text{MHz}$), held at $f_{\text{max}}$ for various length of time, and linearly ramped back to zero at the end of transport. This frequency ramp precisely defines the motional profile of the molecular cloud during the transport. The tunable lens is driven by a homemade constant current driver, controlled by an analog voltage signal. The tunable lens driving waveform is precomputed and scaled to track the molecular cloud position during the transport. For a round trip transport, this sequence is reversed with the relative frequency ramped in the opposite direction.

\section{Transport Performance}

We benchmark transport performance by first studying round trip transports, starting and ending in the MOT chamber. This allows direct extraction of the transport efficiency without requiring the determination of photon collection efficiency or calibration of the camera gain in the downstream chamber (``science cell"). To vary the transport distance, we fix the acceleration and total transport duration and vary the acceleration duration reaching different maximum velocities. We take two $\Lambda$-imaging pictures, each with $20\,\text{ms}$ exposure time, before and after the round trip transport. We then normalize the camera signal of the second image relative to the first image to remove molecule number fluctuation.

\begin{figure}[!htbp]
\subcaptionOverlay{\includegraphics[width=0.5\columnwidth]{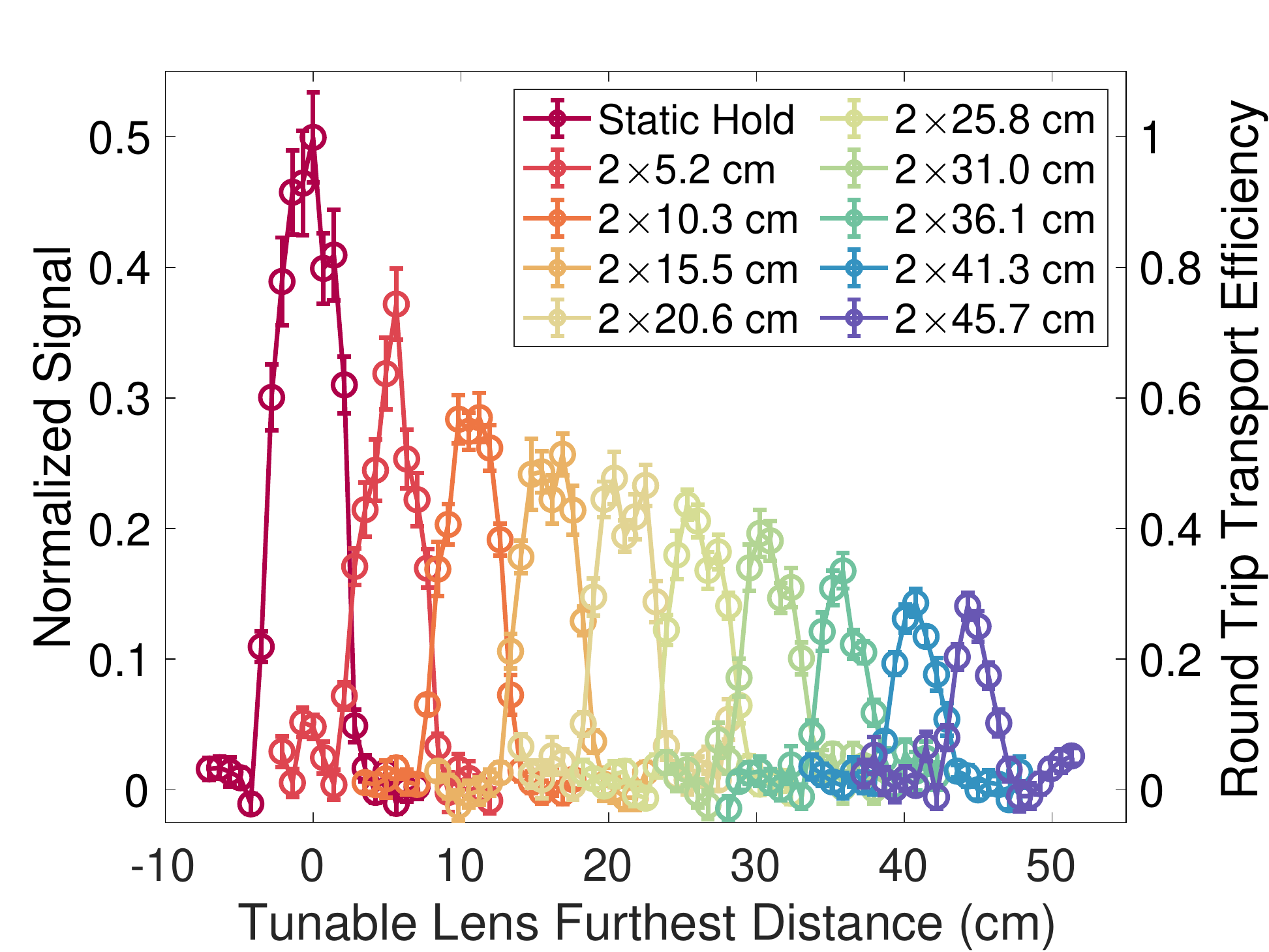}}{(a)}{-2ex,0.5ex}
\subcaptionOverlay{\includegraphics[width=0.5\columnwidth]{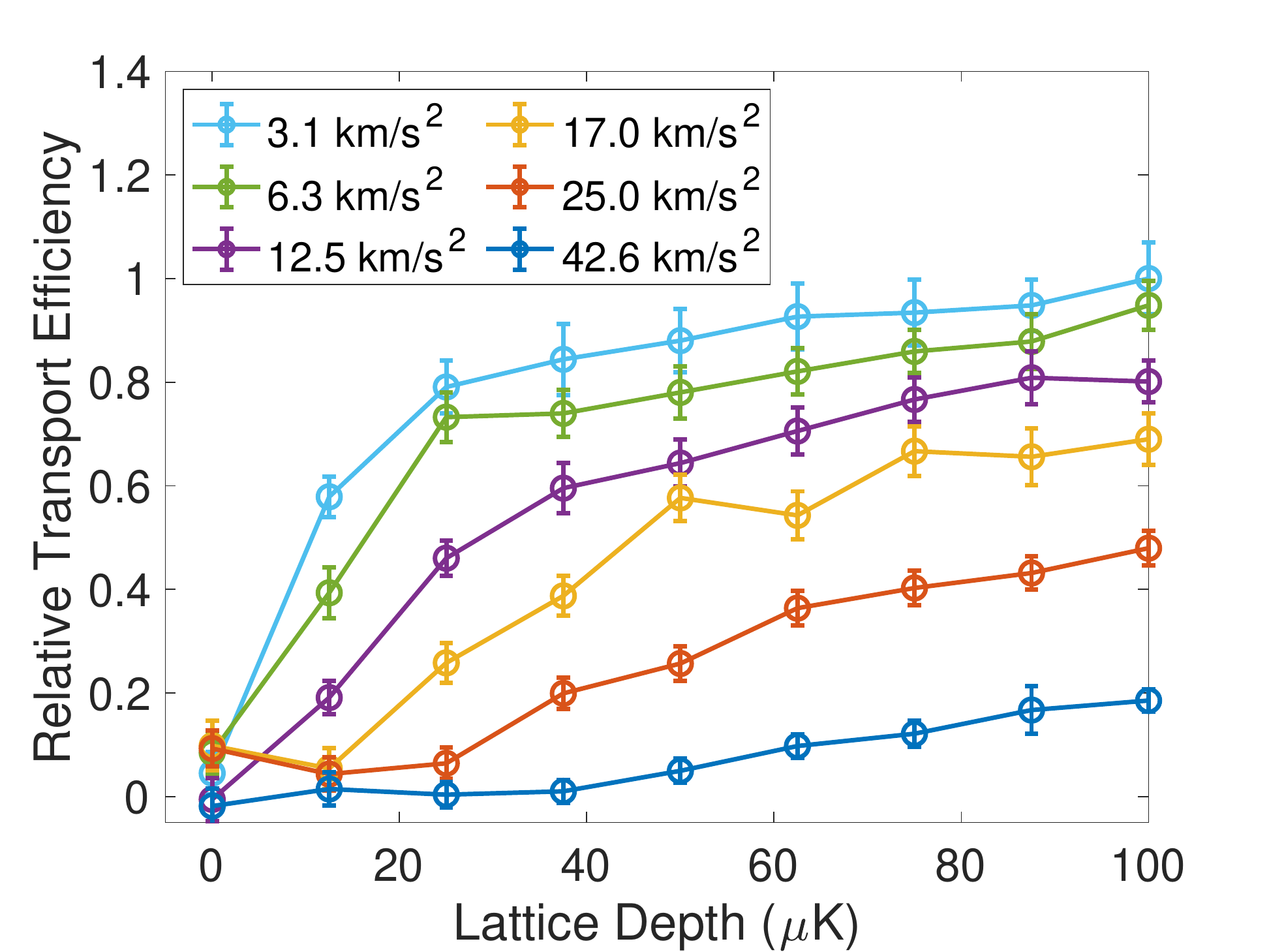}}{(b)}{-2ex,0.5ex}
\caption{(a) Transport efficiency versus the furthest distance reached by the tunable lens focus, taken with different transport distance defined by the lattice motion. Data points on each trace is obtained by scanning the amplitude of the tunable lens driving waveform. The left vertical axis shows the normalized fluorescence signal after transport referenced to the fluorescence signal before the transport. The right vertical axis shows the transport efficiency calculated by normalizing the signal relative to the peak value of static hold condition. The error bars are all referenced to the left vertical axis. All transports shown in this figure are round trip, with the longest distance being 45.7cm to science cell and then back to MOT chamber. The acceleration is $6.3\,\text{km}/\text{s}^2$ and the round trip transport time is $100\,\text{ms}$. (b) Relative transport efficiency versus the lattice depth with different acceleration during acceleration phases. Relative transport efficiency is all normalized relative to the peak efficiency in this data set.}
\label{fig:results}
\end{figure}

Molecule loss is observed with no transport, i.e. holding molecules in a static lattice. As shown in \Fref{fig:results}(a), after holding molecules in a static lattice for $100\,\text{ms}$, the normalized signal drops to 50\%. Most of this loss is caused by the diffusion of molecules out of the low trap frequency ODT during the $\Lambda$-imaging. A small portion of the loss can be assigned to the lattice shaking originated from the technical limitations such as the phase noise picked up in the fibers and residual intensity noise of the laser. As the lattice beam power is being lowered, the lattice trap frequency decreases, and we observe a longer lifetime in the lattice. Therefore, to reduce loss during the constant velocity transport, we find it beneficial to lower the lattice beam power and release molecules back to the ODT. All data shown in \Fref{fig:results} is taken with the lattice beam power ramped to zero during the constant velocity period of the transport.

We scan the lattice depth by varying the lattice beam power while keeping the ODT beam power constant. We measure the transport efficiency at roughly half of the maximum round trip transport distance with different combinations of lattice depth and acceleration as shown in \Fref{fig:results}(b). An accelerating lattice can be viewed as a tilted lattice with reduced depth. For low acceleration, even a reduced depth is enough to hold molecules in the lattice site. The data indicates that when the acceleration is low, the transport efficiency saturates as lattice depth increases. For higher acceleration, the transport efficiency increases gradually as lattice becomes deeper, but the overall transport efficiency is lower. This is expected because increasing the overall lattice depth can compensate for the larger reduction in lattice depth caused by higher acceleration. When the lattice beam is not turned on, molecules are not transported. This demonstrates that our moving lattice can provide a higher axial acceleration than that of a simple ODT.

The highest transport efficiency is reached at an optimal acceleration. High acceleration can increase loss, while too low of an acceleration will increase the time that the lattice has to be turned on, leading to lattice heating loss. The transport efficiency reported in this work is achieved with an acceleration around $6.3\,\text{km}/\text{s}^2$. For one way transport, this accelerates molecules to $9.4\,\text{m/s}$ in $1.5\,\text{ms}$. With a total $50\,\text{ms}$ transport time, molecules are transported $45.7\,\text{cm}$ to the center of science cell.

We measure the temperature of the trapped molecules before and after transport using time-of-flight method. The temperature of the molecules before transport is $32(2)\,\mu\text{K}$. This temperature is higher than $\sim10\,\mu\text{K}$ achieved in the free space, because the tensor AC stark shift from the deep optical trap affects the $\Lambda$-cooling \cite{cheuk2018lambda}. We then apply the same $2\times20.6\,\text{cm}$ round trip transport using $6.3\,\text{km}/\text{s}^2$ acceleration and re-measure the temperature. The temperature of the molecules after transport rises up to $53(4)\,\mu\text{K}$. However, this is still within the capture velocity of $\Lambda$-cooling. By applying cooling again in the science cell, we can recover the lowest trapped molecule temperature.

Using one way transport, we take a second image of the molecular cloud in the science cell. Imaging in the science cell shows a transport efficiency of $48(8)\%$. The slight difference of this efficiency compared to the square root of round trip efficiency ($\sqrt{28\%}\approx53\%$) is due to the the calibration uncertainty of the cameras and the uncertainty of imaging optics collection efficiency between MOT chamber and science cell. Using these transported molecules, we can consistently load optical tweezers with near $50\%$ probability in the science cell.

\section{Conclusion}
In conclusion, we present a hybrid scheme for transporting ultracold laser-cooled molecules over long distances. The combination of a tunable lens ODT and moving lattice allows high speed transport and direct loading of the transport lattice. We also find that the density of the transported molecular cloud is sufficient to reliably load optical tweezers of CaF molecules. With the high numerical aperture optical access available in the science cell, this paves the way for further studies such as Raman sideband cooling of molecules \cite{caldwell2020sideband}, the realization of 2-qubit gate operation between molecules and larger tweezer arrays \cite{ni2018dipolar,hughes2020robust,caldwell2020enhancing,caldwell2021general}. This transport scheme is universal to any atomic or molecular species that can be optically trapped in an optical lattice. It is also possible to upgrade from a tunable-lens only moving ODT transport scheme by adding the large waist lattice beam, whenever high transport velocity is desired.

\section{Acknowledgements}
This work was supported by DOE Quantum System Accelerator, AFOSR, ARO and NSF. SB and SY acknowledge support from the NSF GRFP. LA and SY acknowledge support from the HQI. EC acknowledges support from the NRF of Korea (2021R1C1C1009450, 2020R1A4A1018015, 2021M3H3A1085299). We thank Ni group for lending a 20D version of the tunable lens for initial testing. We thank Julian Léonard for fruitful discussions about technical details regarding tunable lens.

%Supplemental material
\section{Appendix}

\subsection{CaF RF MOT}
CaF molecules are first produced by chemical reaction between laser ablated calcium atoms and sulfur hexafluoride ($\text{SF}_6$) gas in the CBGB source. The CBGB source is cooled to $2.3\,\text{K}$ by a pulse tube cryocooler with a closed cycle liquid helium pot. The molecular beam is then radiatively slowed by a counter-propagating frequency chirped laser pulse resonating with X-B transition of CaF. Vibrational repump lasers addressing $v=1$ and $v=2$ vibrational states in the ground electronic states are sent together along the X-B laser into the MOT chamber. The linearly polarized X-B laser is polarization-switched between two orthogonal polarization states using a Pockels cell to destabilize dark states during the slowing. The $v=1$ vibrational repumps are frequency-broadened by a “white light” EOM\cite{hemmerling2016laser}. This EOM is resonantly driven by a tank circuit with $5\,\text{MHz}$ RF to reach a modulation index over 50. This effectively broadens the laser to $>500\,\text{MHz}$ bandwidth. Both vibrational repumps are then passed through an EOM driven at 25MHz and modulation index of 3.83 to add sidebands for addressing all hyperfine states in $N=1$ rotational state of CaF. $v=3$ vibrational repump laser is sent into the chamber through an auxiliary viewport.

The MOT coils consist of a pair of in-vacuum magnetic coils, forming a resonant tank circuit with variable capacitors. Each tank circuit is individually driven by a homemade $1\,\text{kW}$ RF amplifier at $1\,\text{MHz}$. This configuration generates up to $50\,\text{G/cm}$ RMS magnetic field gradient during the MOT compression phase. This high magnetic field gradient allows compression the MOT to higher density cloud, boosting the loading efficiency into the transport lattice. The RF MOT starts with $10\,\text{G/cm}$ RMS gradient and total intensity of $126\,\text{mW}/\text{cm}^{2}$ for loading. Subsequently, MOT beam total intensity is ramped down to $18\,\text{mW}/\text{cm}^{2}$ in $5\,\text{ms}$, and the magnetic field gradient is ramped up to $50\,\text{G/cm}$ RMS in $10\,\text{ms}$. This compresses the molecular cloud to a Gaussian RMS width of $0.8\,\text{mm}$.

With no helium flow from the CBGB source, the pressure in the MOT chamber is around $3\times10^{-10}\,\text{Torr}$ and the pressure in the science cell region is around $5\times10^{-11}\,\text{Torr}$. With a typical $1\,\text{sccm}$ helium flow rate in the CBGB source and a in-vacuum shutter opening time of 10ms, we observed a typical pressure spike as high as $1\times10^{-8}\,\text{Torr}$ in the MOT chamber and $1\times10^{-9}\,\text{Torr}$ in the science cell region. The helium introduced into the UHV system is quickly pumped away by turbomolecular pumps and ion pumps. The science cell region can be pumped to below $2\times10^{-10}\,\text{Torr}$ in about $500\,\text{ms}$. This performance can be further improved by reducing the size of the differential pumping aperture.

\subsection{Transport lattice setup}
\begin{figure}
\includegraphics[width=\columnwidth]{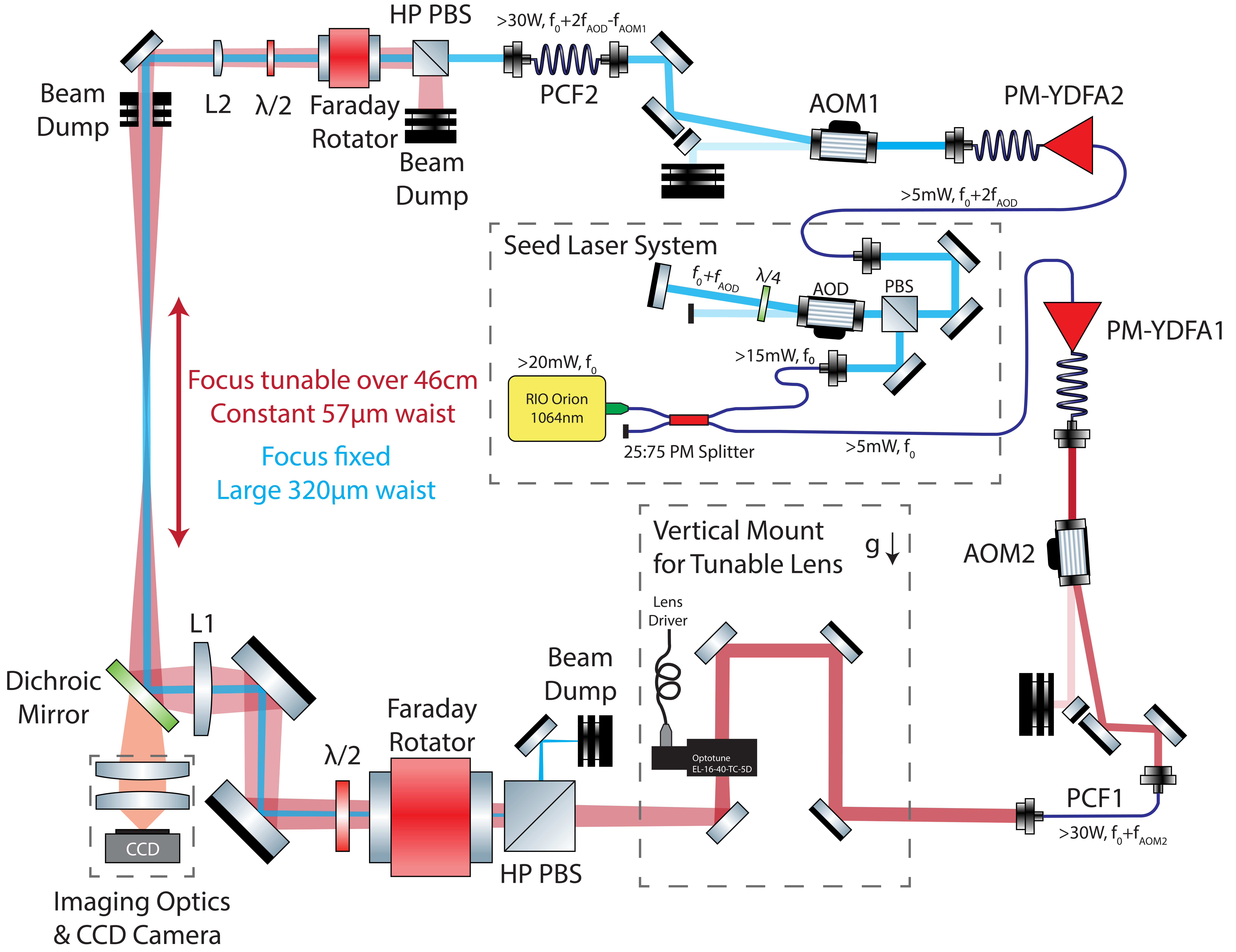}
\caption{Transport lattice setup view from the top. The tunable lens mount section inside the dashed box is vertically mounted on the optical table.}
\label{fig:detail_setup}
\end{figure}

The optical setup of the transport lattice is shown in \Fref{fig:detail_setup}. Trapping lasers for both ODT beam and lattice beam are derived from a low intensity noise narrow linewidth ($<5\,\text{kHz}$) laser module (RIO Orion $1064\,\text{nm}$) emitting over $20\,\text{mW}$ of single frequency $1064\,\text{nm}$ into a PM fiber. This seed laser is split by a 25:75 PM fiber splitter into $5\,\text{mW}$ and $15\,\text{mW}$. $25\%$ of the seed power is directly fed into a $50\,\text{W}$ single frequency PM-YDFA (PM-YDFA1) from Precilasers. The rest of the seed power is free space collimated, sent through a double passed acoustic optic deflector (AOD) setup and then re-coupled into another PM-YDFA of the same model (PM-YDFA2). An acoustic beam steering AOD is chosen here to maintain high diffraction efficiency over wide range frequency range ($f_{\text{AOD}}=60\,\text{MHz}\sim100\,\text{MHz}$, $f_{\text{shift}}=+120\,\text{MHz}\sim+200\,\text{MHz}$). We obtain more than $5\,\text{mW}$ fiber coupled seed power into PM-YDFA2 over the full $2\times40\,\text{MHz}$ tuning range of the double passed AOD. The PM-YDFAs are working in saturated regime where the output power is less sensitive to input seed power fluctuation compared to a non-saturated optical amplifier. We observe no output power fluctuation as we tune over the full $2\times40\,\text{MHz}$ tuning range.

The outputs of both PM-YDFAs are both optically isolated and sent through AOMs used as optical switches and attenuators. The PM-YDFA1 output is frequency shifted down by an AOM driven at $f_{\text{AOM1}}=80\,\text{MHz}$ (AOM1) and then coupled into a photonics crystal fiber (PCF1). The output of this fiber is delivered to the ODT beam section. The PM-YDFA2 output is frequency shifted up by an AOM driven at $f_{\text{AOM2}}=80\,\text{MHz}$ (AOM2) and then coupled into another PCF of the same type (PCF2). The output of PCF2 is delivered to the lattice beam section. We achieve $>30\,\text{W}$ optical power on the experimental table for each beam. The frequency difference between two beams after PCFs is $\Delta f=f_{\text{PCF1}}-f_{\text{PCF2}}=f_{\text{AOM1}}+f_{\text{AOM2}}-2f_{\text{AOD}}$. By driving the double passed AOD at frequency $f_{\text{AOD}}=80\,\text{MHz}$, the frequency difference is zero, corresponding to a static lattice condition. Tuning $f_{AOD}$ above $80\,\text{MHz}$ moves lattice sites towards MOT chamber direction and tuning $f_{\text{AOD}}$ below $80\,\text{MHz}$ moves lattice sites towards science cell direction.

The $f_{\text{AOD}}$, $f_{\text{AOM1}}$ and $f_{\text{AOM2}}$ signal are generated by AD9910 direct digital synthesizer (DDS) based on Analog Devices’s EVAL-AD9910 evaluation board. Both synthesizers are clocked by a shared $1\,\text{GHz}$ low phase noise crystal oscillator (CCSO-914X3-1000) from Crystek. The phase lock loop (PLL) in the AD9910 DDS is bypassed to eliminate the phase noise from PLL’s servo bump. The $f_{\text{AOM1}}$ and $f_{\text{AOM2}}$ signals are single tone $80\,\text{MHz}$ sine wave generated by a single AD9910. The output of this DDS is split into two paths, separately attenuated and amplified to drive AOM1 and AOM2. The driving amplitude into these two AOMs is controlled by intensity stabilization servos that stabilize the output power after each PCF. The frequency swept $f_{\text{AOD}}$ signal is generated by a separate AD9910 DDS utilizing its internal digital ramp generator (DRG). The internal DRG can only be used to sweep frequency for one way transport. To support round trip transport, an Arduino Due microcontroller board is used to reconfigure the ramp limits of DRG in real time to transport molecules back to the MOT chamber.

The $f_{\text{AOD}}$ signal can be also generated by an arbitrary waveform generator (AWG) loaded with frequency modulated waveform for a specific lattice moving velocity profile. To do so, the AWG loops through a static frequency waveform to generate a fixed frequency sine wave for lattice loading. Then, the AWG will jump to the transport waveform once triggered by the experimental sequence. This guarantees constant seed laser power into the fiber amplifier and prevents damage of the fiber amplifier from losing seed input. Our initial setup used a Tektronix AWG70001A to generate $f_{\text{AOD}}$ where we saw no difference in transport efficiency compared to AD9910 DDS.

One technical note that we would like to point out is that the high power lattice beams need to be properly dumped after passing through the vacuum system. Unlike a retro-reflected optical lattice setup, in our setup, the main focusing lens L1 focuses down the counter-propagating lattice beam to a small spot onto the tunable lens. This high intensity focal spot can easily exceed the damage threshold of the tunable lens. Thus, a high-power large-aperture optical isolator must be installed between L1 and the tunable lens.

\subsection{Optimization of tunable lens driving waveform}

\begin{figure}[H]
\subcaptionOverlay{\includegraphics[width=\columnwidth]{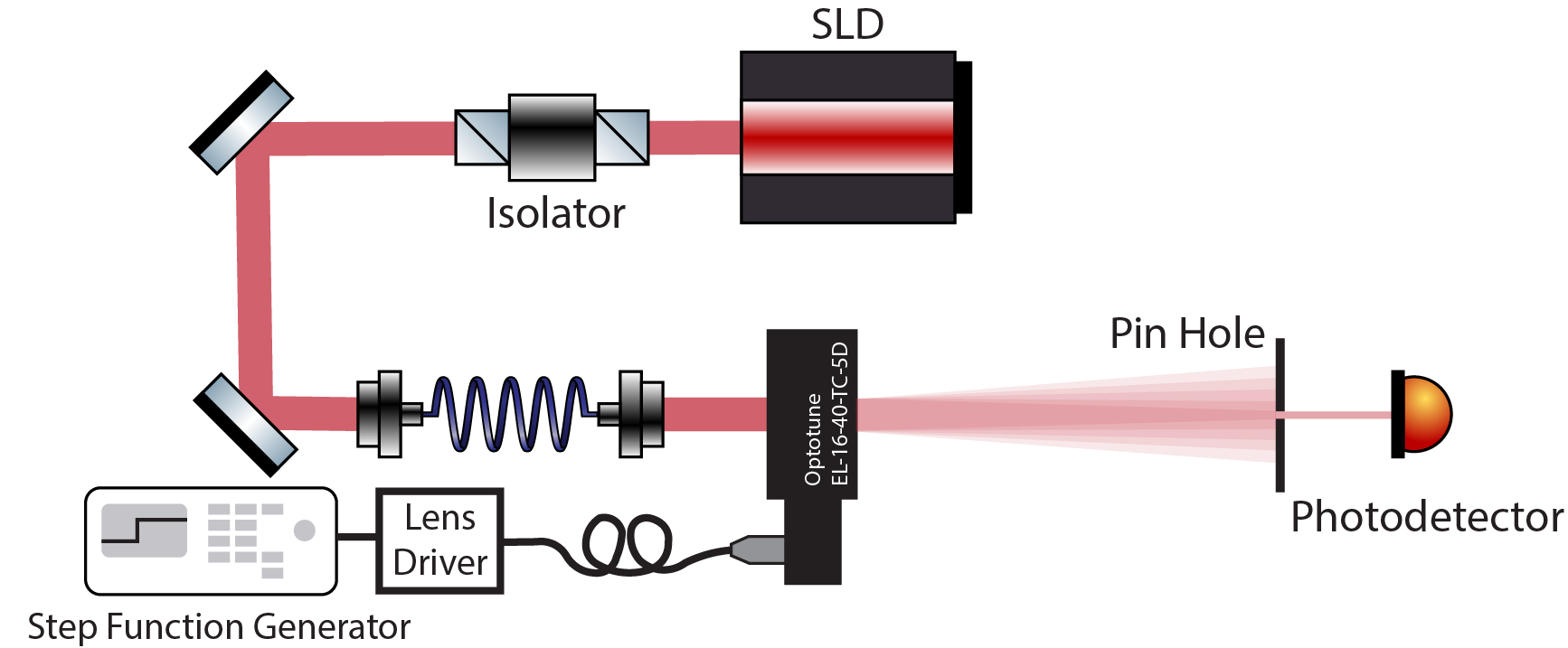}}{(a)}{-2ex,1ex}
\subcaptionOverlay{\includegraphics[width=0.5\columnwidth]{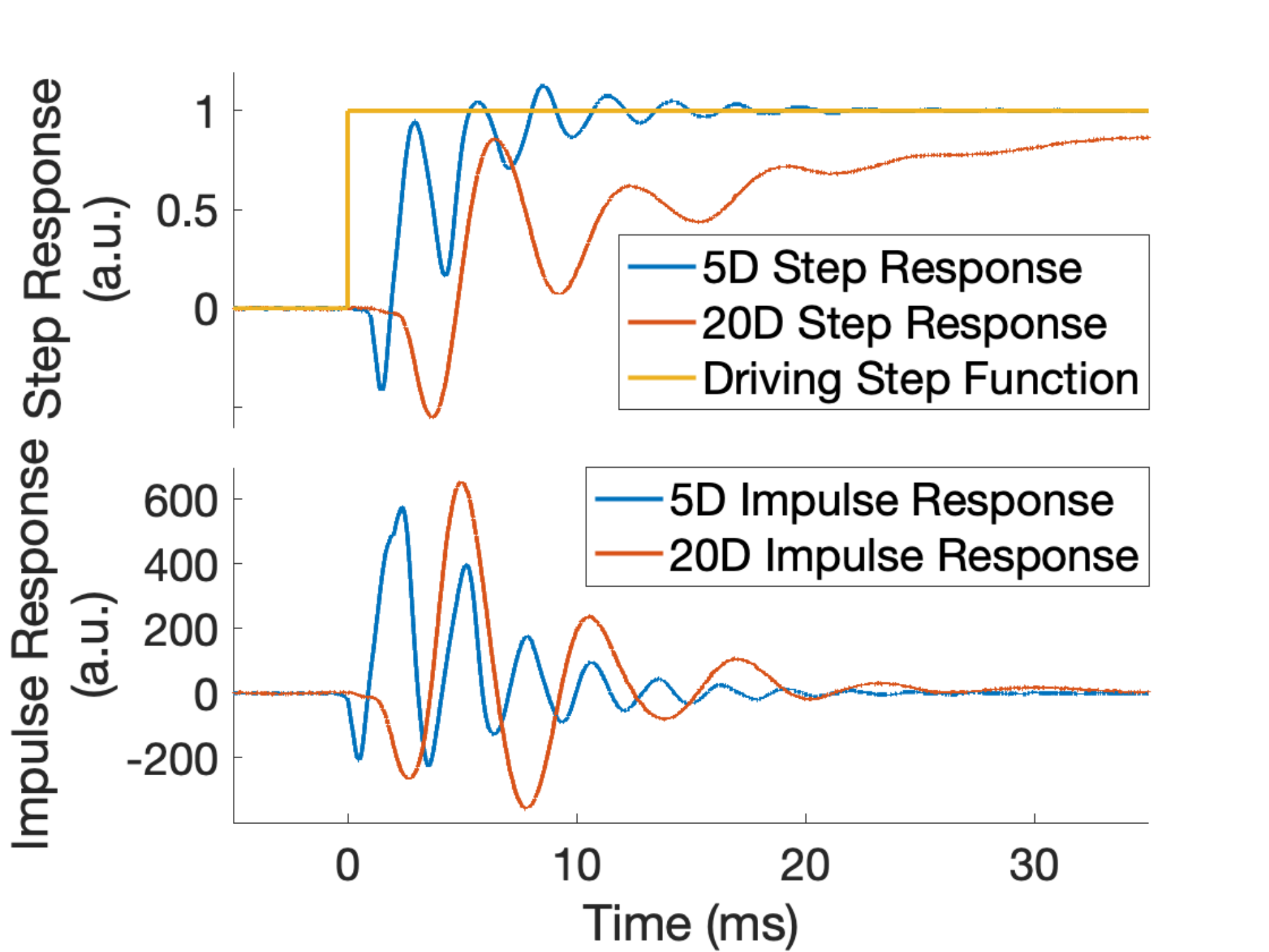}}{(b)}{-2ex,1ex}
\subcaptionOverlay{\includegraphics[width=0.5\columnwidth]{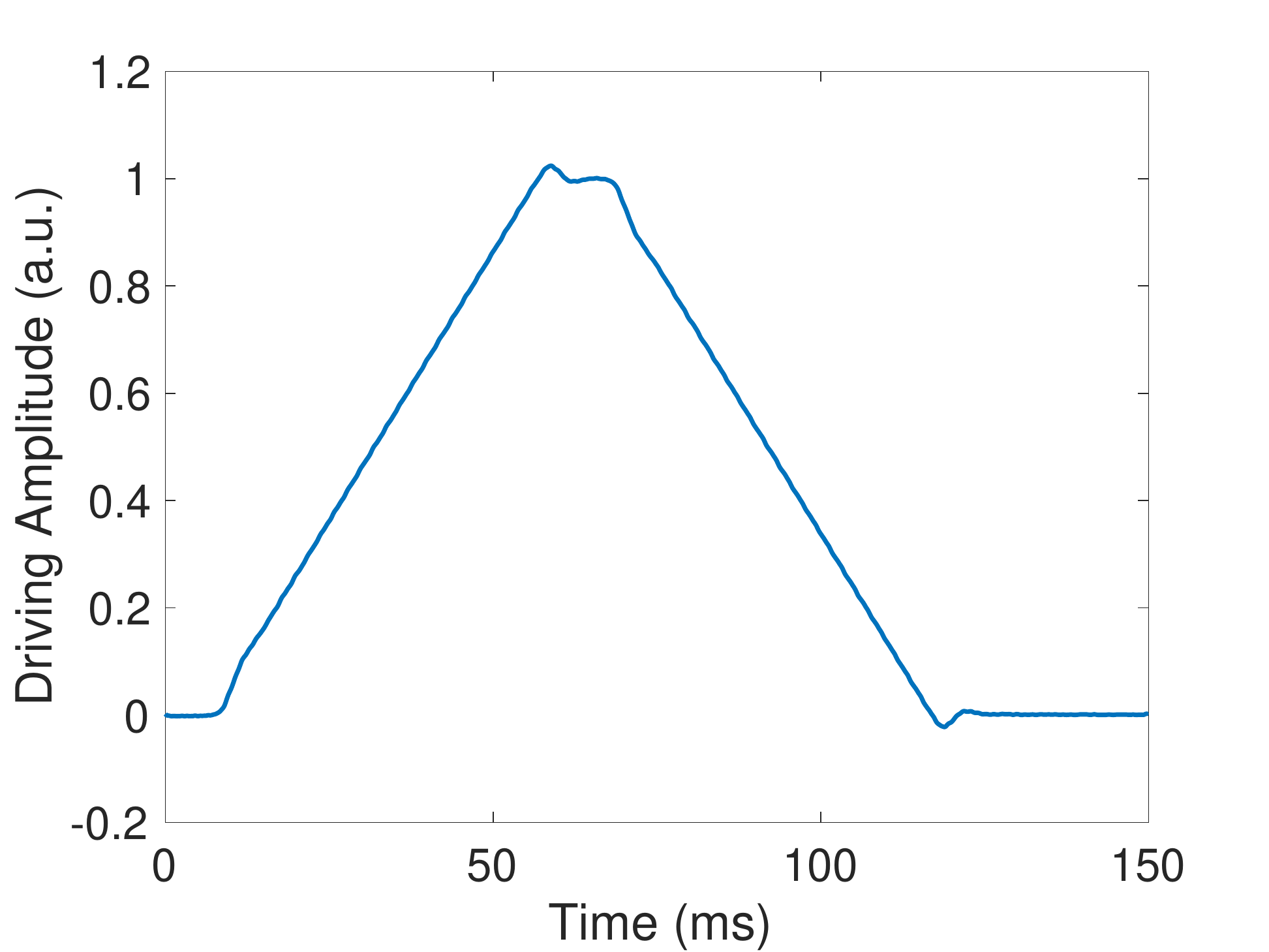}}{(c)}{-2ex,1ex}
\caption{(a) Impulse response measurement setup (b) Measured step response and impulse response functions of 20D and 5D lens. (c) Optimized driving waveform for a simple linear ramp as a target trajectory. }
\label{fig:tunable_lens_response}
\end{figure}

We model the tunable lens as a linear time-invariant (LTI) system. An LTI system can be described by its impulse response function. We measure the impulse response function by applying a step function current waveform to the tunable lens. The response signal is measured by a photodiode behind a pinhole placed on the optical axis of the tunable lens (\Fref{fig:tunable_lens_response}(a)). We observe some interference fringes when using a laser to illuminate. We believe this originates from surface reflections of the cover glass and tunable lens membranes that appear as fluctuations on the photodiode signal. This fluctuation is eliminated by using a single mode fiber coupled super luminescent diode (Q-Photonics QSDM-790-30) as an incoherent light source. The photodiode signal is recorded by a 16bit digitizer. For a gaussian beam with a $1/e^2$ diameter of $w_0$ as input, the intensity at the center of the beam follows $P=\frac{\pi w^2 I}{2}$, where $P$ is the total power in the beam and $w$ is the $1/e^2$ beam radius at the pinhole plane. The distance $d$ from the ODT focus to L1 satisfies $\frac{1}{d}=\frac{1}{f_{L1}}-\frac{1}{f_{L1}w/(w-w_{0})}$. It is easy to derive $d=\frac{f_{L1}w}{w0}=\frac{f_{L1}\sqrt{\frac{2P}{I}}}{w0}\propto\frac{1}{\sqrt{I}}$. In reality, because of slight misalignment between the beam and tunable lens's optical axis, the photodiode signal deviates from the formula when $w$ is small (lens is slightly focusing instead of diverging the beam). We instead use an interpolated function between photodetector output versus measured lens position to linearize the tunable lens response. The tunable lens position signal is then differentiated to derive the impulse response function $h(t)$. Here we provide two examples of the impulse response function from our 20 diopters (20D) range and 5 diopters (5D) range models of the lens in \Fref{fig:tunable_lens_response}(b), showing very different resonant and damping characteristics. The impulse response function is also different from lens to lens. Thus, each new lens needs to be tested for its unique impulse response function before the optimization algorithm can be applied. The 5D lens with a stiffer membrane is chosen for this experiment, which has a higher resonant frequency and can be driven faster. 

By taking the convolution of an arbitrary driving waveform $x(t)$ with the measured impulse response function, we can simulate the tunable lens position response $y(t)=x(t)*h(t)$ on a computer, where $*$ denotes discrete convolution. We test the difference between a simulated response to the target response waveform $y_{0}(t)$ using a customized cost function $c(x(t))=x(t)*h(t)-y_{0}(t)+k\int{|x(t')-\int_{t=t'-\tau}^{t'+\tau}{x(t)}|^2 dt'}$. The last term is the penalty term added in the cost function to suppress the high frequency noise components in the optimized waveform, in which the variable $k$ is the weight of the penalty term and $\tau$ is a low-pass filter time constant. We then use the CVX convex optimizer package in MATLAB to minimize this cost function to find the optimized driving waveform \cite{cvx,gb08}. The typical optimized driving waveform used in this experiment is shown in \Fref{fig:tunable_lens_response}(c).

\begin{figure}[!htbp]
\includegraphics[width=\columnwidth]{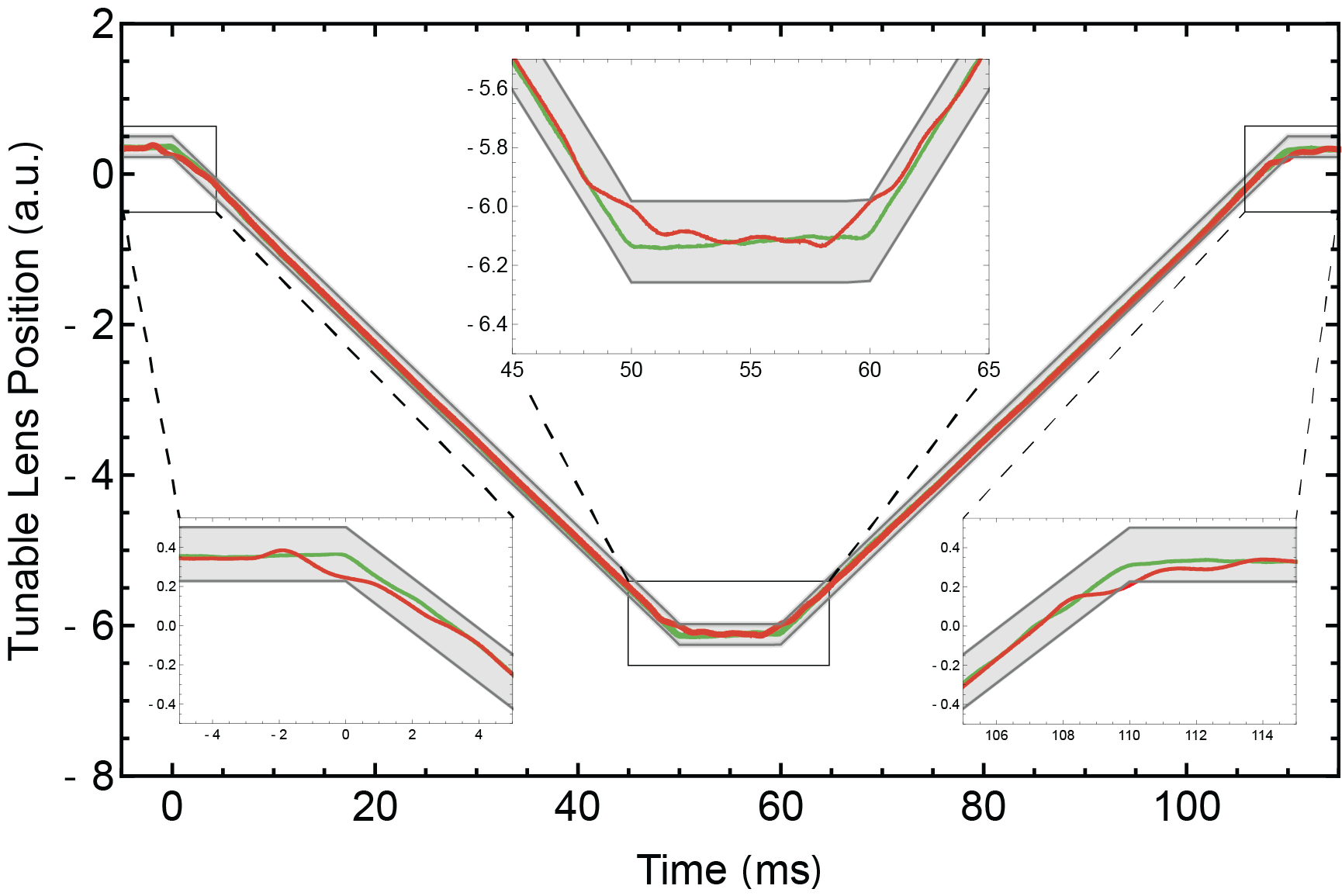}
\caption{Lens response driven with a simple linear ramp (red) versus an optimized waveform (green). The gray area is the Rayleigh range of the ODT along the target trajectory. Key moments during lattice acceleration and deceleration are zoomed in for clarity.}
\label{fig:tunable_lens_optimized}
\end{figure}

In \Fref{fig:tunable_lens_optimized}, we show the distorted response of the tunable lens when driven with a simple linear ramp signal. Driving with an optimized waveform shows no distortion near the beginning and the end of the motion. During the constant velocity transport, there is very small difference between the two waveforms. We find that using the optimized waveform is crucial in achieving high efficiency in our fast transport scheme. We see no molecule transported at full distance when using the simple ramp signal. This can be explained by the distortion caused by higher resonant modes of the lens membrane or large deviation from target trajectory during the acceleration period.

\subsection{Molecule number calibration}
\begin{figure}[H]
\subcaptionOverlay{\includegraphics[width=0.5\columnwidth]{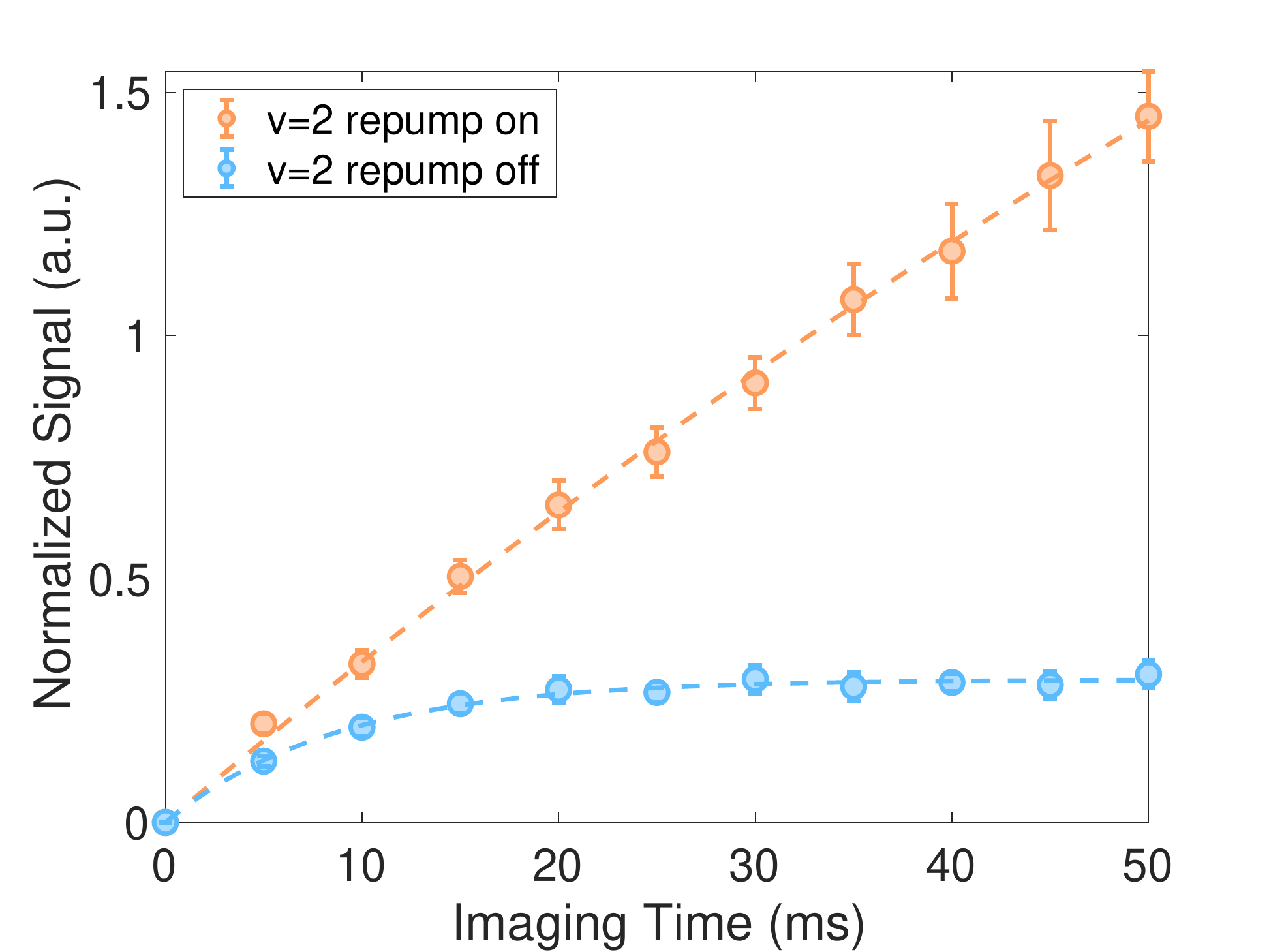}}{(a)}{-2ex,0.5ex}\subcaptionOverlay{\includegraphics[width=0.5\columnwidth]{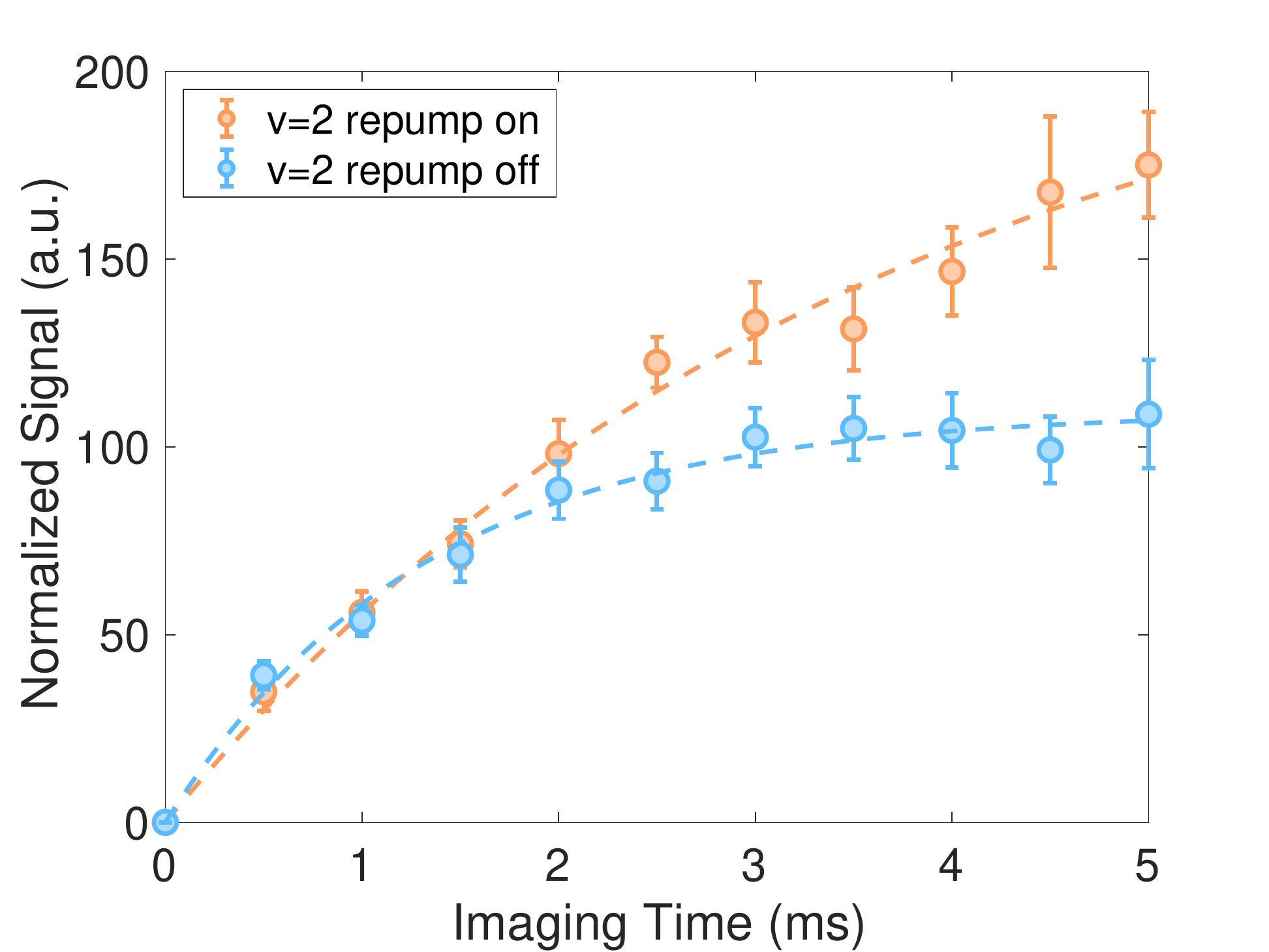}}{(b)}{-2ex,0.5ex}
\caption{(a) v=2 repump on/off during lambda imaging in the MOT chamber. (b) v=2 repump on/off during resonant imaging in the science cell.}
\label{fig:imaging_calibration}
\end{figure}

The molecule number is measured by collecting $606\,\text{nm}$ fluorescence on a camera during $\Lambda$-cooling or resonant imaging. By turning off $v=2$ repump laser, we can set a limit of the average photon number scattered by each molecule. By scanning the exposure time of the camera, we see the total number of scattered photons saturates. This saturation corresponds to that almost all molecules are pumped into the $v=2$ vibrational state. We show the signal received by camera for fluorescence in both MOT chamber and science cell with and without $v=2$ repump laser in \Fref{fig:imaging_calibration}, where the photon scattering saturation level can be easily fitted to an exponential model. Note the signals here are relative values normalized against the first image taken in the MOT chamber before transport in order to remove the fluctuation in molecule number that is caused by YAG ablation fluctuation in CBGB source. By turning on $v=2$ repump laser during the imaging, we can see the signal does not saturate as exposure time increases.

To get the actual number of molecules in the trap, we need to know the relation between total photon scattered versus the signal on camera. We calibrate the gain of our camera by sending a 606nm laser beam with calibrated power onto the camera through all imaging optics. This ensures that we take into account the loss in imaging optics. We use the geometric numerical aperture to derive the photon collection efficiency. This efficiency is then used to derive the total number of photons scattered from the molecular cloud into full solid angle. The imaging beam size, power, imaging optics and cameras are all different between MOT chamber and science cell. Thus, all the calibration needs to be carried out separately on each side.

\subsection{Time-of-flight (ToF) measurement}

\begin{figure}[H]
\subcaptionOverlay{\includegraphics[width=0.5\columnwidth]{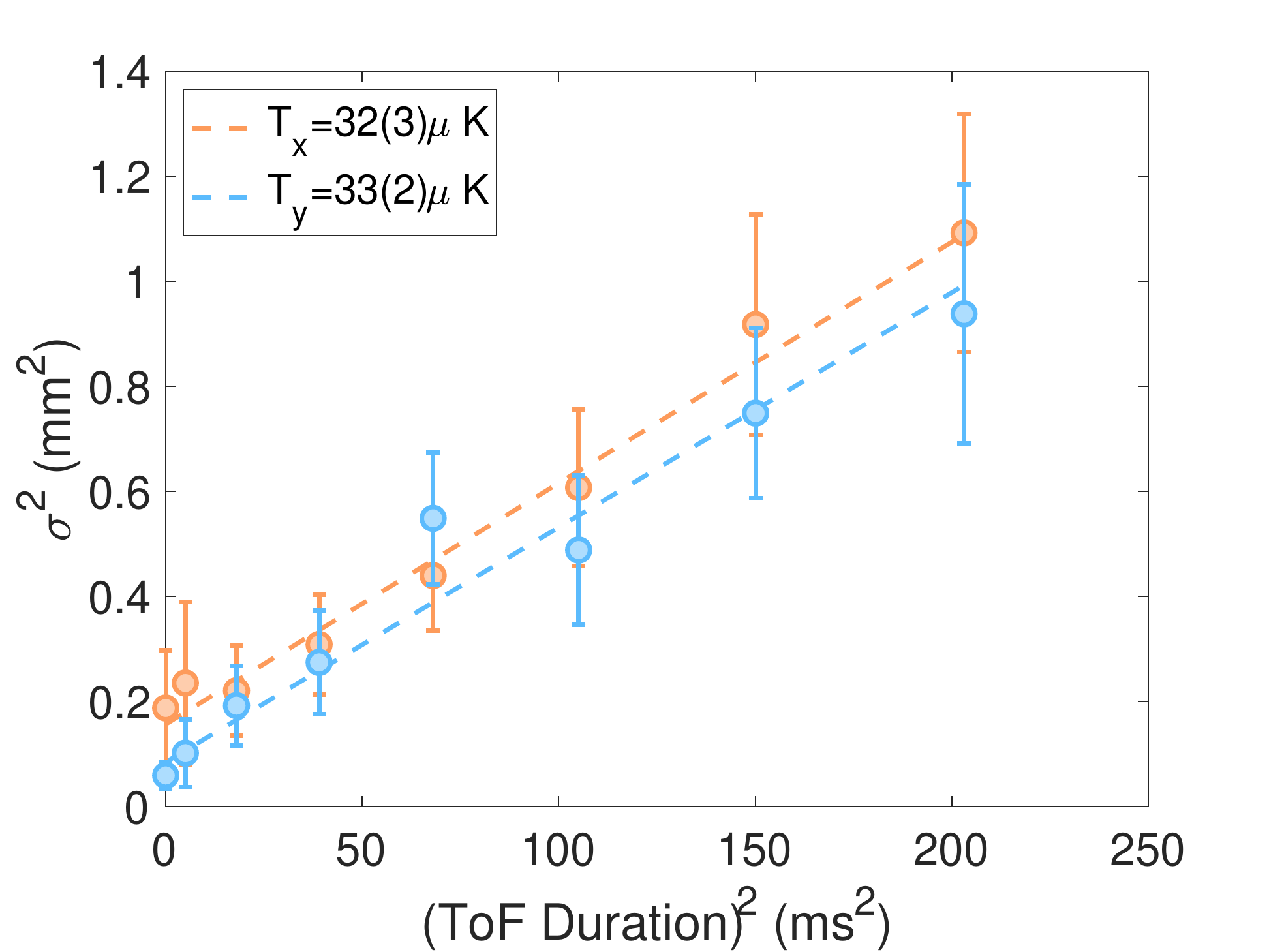}}{(a)}{-2ex,0.5ex}
\subcaptionOverlay{\includegraphics[width=0.5\columnwidth]{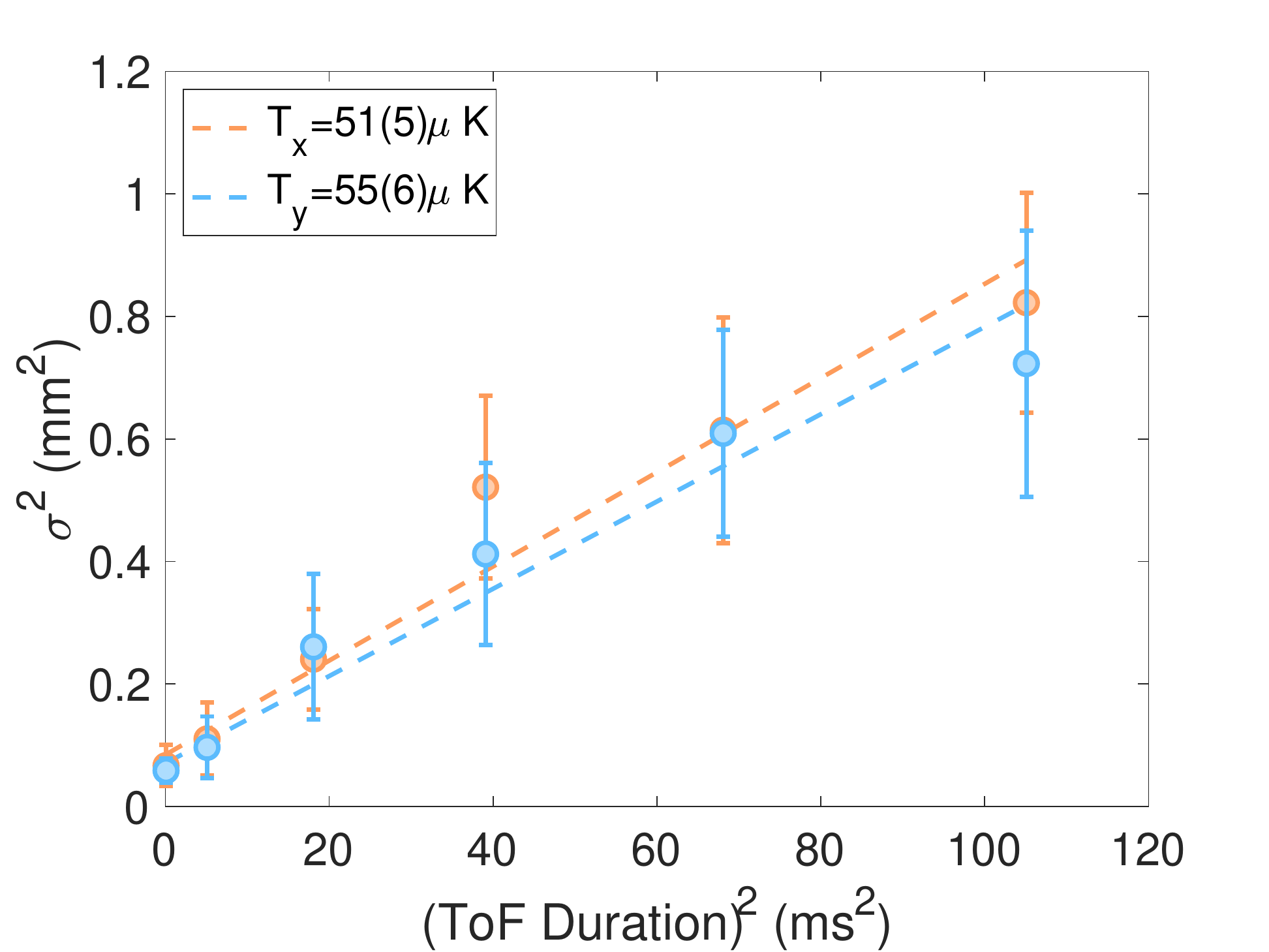}}{(b)}{-2ex,0.5ex}
\caption{(a) ToF measurement of trapped molecules after $100\,\text{ms}$ long hold time. (b) ToF measurement of trapped molecules after $100\,\text{ms}$ round trip transport.}
\label{fig:ToF}
\end{figure}

We turn off the trap and wait for various lengths of time. We then turn on a $1\,\text{ms}$ long resonant imaging pulse to image the expanded molecular cloud. The ToF measurement data and fitted temperature on the two axes are shown in \Fref{fig:ToF}. 

\section*{References}
\bibliographystyle{iopart-num}
\bibliography{transport} 
\end{document}